# Melting of antiferromagnetic order in $La_{0.5}Sr_{1.5}MnO_4$ probed with ultrafast resonant soft X-ray diffraction

H. Ehrke[1,2], R.I. Tobey[1,3], S. Wall[1,4], S. A. Cavill[2], M. Först[3], V. Khanna[3], Th. Garl[3],
N. Stojanovic[5], D. Prabhakaran[1], A.T. Boothroyd[1], M. Gensch[6], A. Mirone[7],
P. Reutler[8], A. Revcolevschi[8], S. S. Dhesi[2]*, A. Cavalleri[1,3]*

[1] *Department of Physics, Clarendon Laboratory, University of Oxford, United Kingdom*
[2] *Diamond Light Source, Harwell Science and Innovation Campus, Chilton, Didcot, UK*
[3] *Max Planck Research Department for Structural Dynamics, University of Hamburg-CFEL, Germany*
[4] *Fritz Haber Institut of the Max Planck Society, Berlin, Germany*
[5] *HASYLAB at DESY, Notkestrasse 85, 22607 Hamburg, Germany*
[6] *Helmoltz Zentrum Berlin, Germany*
[7] *European Synchrotron Radiation Facility, BP 220, F-38043 Grenoble Cedex, France*
[8] *Laboratoire de Physico-Chimie de l'Etat Solide, URM8182, Université Paris Sud, Orsay, France*

## Abstract

We show how ultrafast resonant soft x-ray diffraction can separately probe the photo-induced dynamics of spin and orbital orders in $La_{0.5}Sr_{1.5}MnO_4$. Ultrafast melting of CE antiferromagnetic spin order is evidenced by the disappearance of a (¼ ¼ ½) diffraction peak. On the other hand the (¼ ¼ 0) peak, reflecting orbital order, is only partially reduced. Cluster calculations aid our interpretation by considering different magneticalyy ordered states accessible after photo-excitation. Nonthermal coupling between light and magnetism emerges as a primary aspect of photo-induced phase transitions in manganites.

Electronic order on nanometre length scales is a common feature of strongly-correlated electron systems. Charges, spins and orbitals form regular patterns to minimise the total potential energy of the system, whilst competing with the opposing tendency of electrons to become itinerant to minimize their kinetic energy. Large-bandwidth manganites like $La_{1-x}Sr_xMnO_3$ are a typical example of such competition, with the antiferromagnetic Mott insulating parent compound ($LaMnO_3$) transforming into double-exchange metallic ferromagnets for hole-doping between $0.2 \leq x \leq 0.5$, and again into complex insulating phases, with charge and orbital ordering, for commensurate doping levels (*i.e.* $x = ½$). Doped manganites can also be controlled reversibly with external stimuli, such as magnetic fields[1,2,3], pressure[4], electric currents[5], x-rays[6], and visible[7] or mid infrared radiation[8,9]. In particular, photo-excitation[10] transfers charge across semicovalent bonds, drastically perturbing spin and orbital orders[11]. Light may then be used in compounds like magnetoresistive manganites to control magnetism on nanometre lengthscales and ultrafast timescales.

Despite the great interest in the photo-induced phenomena, a comprehensive understanding of the underlying physics is missing. This is particularly elusive because charge, spin and orbital arrangements are interdependent degrees of freedom, evolving on theultrafast timescale and on nanometer lengthscales, and are thus difficult to disentangle as they evolve in time[12]. Transient orders can only be probed with ultrafast techniques sensitive to nanometre-scale modulations of charge and spin densities. These modulations are typically small perturbations on the total charge at each atomic site, and are not accessible with any of the time-resolved optical[13] or x-ray[14,15] probes used to date.

Statically, electronic order can be probed with Soft X-ray Resonant Diffraction (SXRD), which is directly sensitive to the relevant electronic states close to the Fermi level by using photon energies resonant with the $2p \rightarrow 3d$ dipole transitions (Mn $L_{2,3}$ edges). Furthermore, the energy dependence of the SXRD intensity can be used to understand the competing interactions leading to spin and orbital ordering[16,17,18,19,20,21].

Here, we have extended SXRD to the ultrafast timescale, and separated ultrafast spin and orbital dynamics in the single-layer, half-doped manganite $La_{0.5}Sr_{1.5}MnO_4$. This is achieved by detecting time dependent diffraction at two different scattering peaks, which reflect spin and orbital order periodicities independently. Our results demonstrate that light excitation completely removes spin order in a non-thermal manner, and only weakly perturbs orbital order.

In our experiments, the base temperature of $La_{0.5}Sr_{1.5}MnO_4$ was held at 25 K, below charge/orbital ordering ($T_{CO-OO}$=220 K) and Néel temperatures ($T_N$ = 110 K). In this CE antiferromagnetic phase, charge, spins and orbitals form a characteristic pattern, well visualized as a set of *antiferromagnetically*-coupled ferromagnetic "zig-zag" chains of $3x^2-r^2$ ($3y^2-r^2$) orbitals at $Mn^{3+}$ sites[22,23].

Trains of 100-femtosecond, 800-nm-wavelength optical pulses at a repetition rate of 20 kHz were synchronized to soft x-ray pulses from a synchrotron storage ring at beamline I06 of the Diamond Light Source. Optical-pump – SXRD-probe experiments measured time dependent (¼ ¼ ½) and (¼ ¼ 0) diffraction peaks, which reflected time dependent spin and orbital ordering, respectively[16,17,24]. These soft x-ray scattering peaks were resolved using photon counting electronics, despite gating of only those pulses synchronous with the laser repetition rate, *i.e.* less than one probe pulse every 20000. The static SXRD energy dependences, as well as those measured when the pump-probe time delays were negative, were in good agreement with those reported in the literature[16,17,21].

$La_{0.5}Sr_{1.5}MnO_4$ was photo excited with light polarized in the *ab* plane, triggering large changes

in SXRD. Figures 1a and 1b show the photon-energy dependence of the (¼ ¼ ½) and (¼ ¼ 0) diffraction peaks 200ps after laser excitation. For all excitation fluences above 5 mJ/cm$^2$ the (¼ ¼ ½) peak entirely vanishes, leaving only a weak signal due to fluorescence. In contrast, the photo-induced reduction in the (¼ ¼ 0) scattering is less than 25%, saturating at the same 5 mJ/cm$^2$ as the (¼ ¼ ½) peak.

Figure 2a show time-dependent scans, recorded at 641.5eV for the (¼ ¼ ½) and at 640.25eV for the (¼ ¼ 0) peak, respectively. The scattered intensities were reduced within the time resolution of the x-ray probe (~50ps) and did not recover within the time window available using a mechanical delay stage (~500ps).

We first note that spin and orbital orders also rearrange differently upon static heating, and that if the temperature increase were to exceed the Néel temperature ($T_N$=110 K), the (¼ ¼ ½) reflection would vanish, whilst the (¼ ¼ 0) reflection would only gradually decrease up to $T_{CO/OO}$=220 K [22]. However, this does not explain our observations. Firstly, a conservative estimate of the temperature increase show that the changes show that the sample temperature remains well below $T_N$, and the changes are nonthermal[25]. Secondly, not only is the maximum possible temperature increase well below $T_N$, but the dependence of the diffraction intensities on laser fluence (see figure 2b) scales very differently than expected for heating. In figure 2c, the measured temperature dependent diffraction is reported for comparison, with the horizontal scales aligned to the fluence dependent graph of figure 2b by calculating sample heating for each fluence. Particularly striking is the weak excitation regime, in which we measure a significant photo-induced drop in diffraction intensity already for 0.66 mJ/cm$^2$, whilst the thermal response of the magnetic peak is temperature independent.

Shorter time delays were probed by using the low-α mode of the Diamond storage ring, with a time resolution of 9 ps. The (¼ ¼ ½) diffraction disappears promptly already on this timescale (Figure 3a and 3b). Longer timescales, up to one microsecond, were probed by electronically controlling the pump-probe separation out to microsecond delays. The diffraction peak returns back to towards the static value with at least three timescales, two of which were fitted with exponentials time constants of 10ns and 110 ns. A much longer timescale is visible as a plateau at about 15% of the diffracted-intensity loss. We stress that the long timescales needed to reach the ground state cannot be simply related to heat diffusion, which is estimated to be only few nanoseconds. Rather, we note that the photo-induced phase is protected by a kinetic barrier, and that return to the ground state is only possible after a complex relaxation path, involving nucleation andgrowth.

Our interpretation proceeds along the following lines. The initial excitation, not accessible on the timescale probed here, involves charge transfer of $e_g$ electrons between $Mn^{3+}$ and the $Mn^{4+}$ sites[26]. As discussed elsewhere[27], photo-excitation of semicovalent bonds affects super-exchange coupling, and is likely to trigger rearrangements in spin ordering, already giving a foretaste of the underlying dynamic physics of transient electronic order. Whilst these first events are beyond the resolution of the present study, we can analyse the metastable state measured at $\tau > 10$ ps on quantitative grounds.

Cluster calculations were performed with the $3d$ states of a central active $Mn^{3+}$ site coupled to the neighboring oxygen $p$-orbitals by a hopping term, modulated by Slater-Koster parameters ref 19. In turn, the in-plane oxygen $p$-orbitals were coupled to the $3d$ states of the neighboring $Mn^{4+}$ sites. This model was used here to explore the nature of several possible excited states. Initially, we considered a phase with spin disordering of all the $Mn^{3+}$ ions. This consisted of a superposition between different $|S|=2$ states, *i.e.* maintaining the same high-spin configuration of the ground state, but with $S_z$ taking all five possible values $S_z=-2,-1,0,1,2$ at different sites. In absence of a thermodynamic description of the product state of the solid, this state is expected to reflect a paramagnetic phase.

A second possible state was also considered, in which the spin alignment was made ferromagnetic on all sites, along the lines of what one may expect for enhanced charge itinerancy. We note that the possibility of a photo-induced ferromagnetic state has also been raised by recent measurements that measured ultrafast Magneto-optical Kerr effect in related compounds[28].

In both cases, we left the Jahn Teller distortion unperturbed, considering that excitations would leave the average charge occupancy unpertubed and would affect primarily short-range correlations. Both paramagnetic and ferromagnetic states have an average excess energy of few tens of meV, and are accessible after the photo-excited system has thermalized[29].

To compare these cluster simulations to our experimental results, we calculated the resonant scattering intensities from either of these transient magnetic states, using the same procedure discussed in reference 19. Figure 4 shows the calculated energy dependence for the (¼ ¼ 0) diffraction peaks for the ground state, as well as for the spin disordered and ferromagnetic states. The (¼ ¼ ½) diffraction intensity was also calculated, although for both paramagnetic and ferromagnetic orders the antiferromagnetic scattering was identically zero.

The calculated (¼ ¼ 0) scattering, reflecting the OO, remained high in intensity at all energies. This is likely related to the existence of a Jahn Teller distortion in the calculation for the excited state. Indeed, evidence for a robust Jahn-Teller orbital ordering comes also from the energy dependence of the experimental (¼ ¼ 0) diffraction intensity shown in figure 1(b), in which the diffracted intensity is reduced more at the two lower energy Mn $L_3$ and $L_2$ edge peaks [21]. In addition, we note that multiplet calculations in a crystal field for mixed spin canted phases in the *ab* plane also show considerable changes in the (¼ ¼ 0) peak spectral profile confirming that the change in intensity of the (¼ ¼ 0) spectral profile is due to spin misorientation[30]

We find that both states (spin disordered and ferromagnetic) give rise to similar *integrated* scattering intensities for the (¼ ¼ 0) peak, and that, based on the available data, we cannot determine if the long-lived photoexcited state is paramagnetic or ferromagnetic. In the future, time resolved x-ray magnetic circular dichroism[31] may provide us with an answer, detecting any net magnetization that develops.

We note that the long lifetimes needed for relaxation are consistent with either of these magnetic states. The reformation of the CE phase, too slow to be understood as thermal diffusion[32], occurs across a kinetic barrier and requires nucleation and domain propagation to be complete. This is schematically depicted in a caricature of the free energy landscape in figure 3c. The system starts in the CE antiferromagnetic ground state and is driven nonthermally into a metastable phase, before relaxing back at long time delays.

The use of time-resolved Soft X-ray Resonant Diffraction to separately probe different contributions to electronic ordering is demonstrated here. In $La_{0.5}Sr_{1.5}MnO_4$, photoexcitation removes spin ordering, whilst the in-plane Jahn-Teller distortion and resulting orbital ordering remains largely unaffected. In the future, measurements at shorter timescales, likely using Free Electron Lasers, will make it possible to distinguish between different contributions in the time domain, and will also access coherences at extreme timescales, including lattice[27], orbital[12] or charge[33] degrees of freedom.

**Acknowledgements:** We would like to thank R. A. Mott for expert technical assistance.
.

**FIGURE CAPTIONS**

**Figure 1**. **(a)** Energy dependence of the (¼ ¼ ½) diffraction peak before photoexcitation (open circles) and 200ps after photoexcitation (red circles). **(b)** Energy dependence of the (¼ ¼ 0) diffraction peak before photoexcitation (open circles) and 200ps after photoexcitation (red circles).

**Figure 2. (a)** Time dependent resonant soft x-ray diffraction at the (¼ ¼ ½) magnetic peak at 640.25 eV, measured with a laser fluence of 5.5 mJ/cm$^2$ (colsed circles) and (¼ ¼ 0) OO diffraction peak at 640.25 eV (closed squares), measured with a laser fluence of 10 mJ/cm$^2$. The red lines are guides to the eye. **(b)** Fluence dependence of the (¼ ¼ ½) and (¼ ¼ 0) diffraction peaks 200 ps after photoexcitation. **(c)** Temperature dependent (¼ ¼ ½) signal, measured statically. Fluence and temperature dependences are aligned by calculating the sample temperature increase for each irradiation fluence.

**Figure 3. (a)** Time-dependent (¼ ¼ ½) diffraction peak at 640.25 eV measured in low-$\alpha$ mode. Red curve: fit of the decay using an error function with a FWHM of 10 ps. **(b)** Energy dependence of the (¼ ¼ ½) magnetic diffraction peak before photoexcitation (open circles with line) and 20ps after photoexcitation (closed red circles with line) **(c)** Decay of the photo induced state (closed circles), together with a double exponential fit (red line). The schematic diagrams for the Free energy ladscape depict the existence of a CE ground state (left side) and of a metastable state (paramagnetic or ferromagnetic) protected by a kinetic barrier.

**Figure 4. Top:** Calculated energy dependence of the (¼ ¼ ½) diffraction peak for the ground state (black) and for two types of metastable states. One for ferromagnetic alignment (red) and a second state for a spin disordered state (blue). **Bottom:** Calculated energy dependence of the (¼ ¼ 0) diffraction peak for the ground state (black) and for two types of metastable states. One for ferromagnetic alignment (red) and a second state for a spin disordered state (blue).

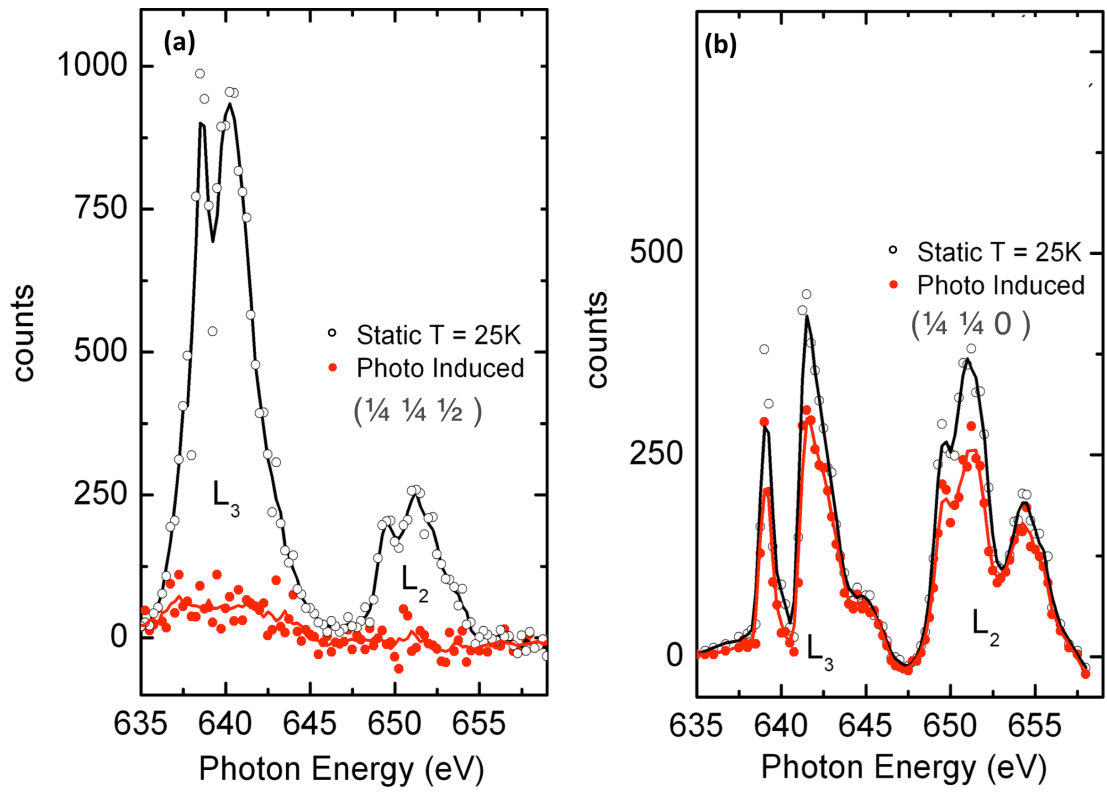

**Figure 1**

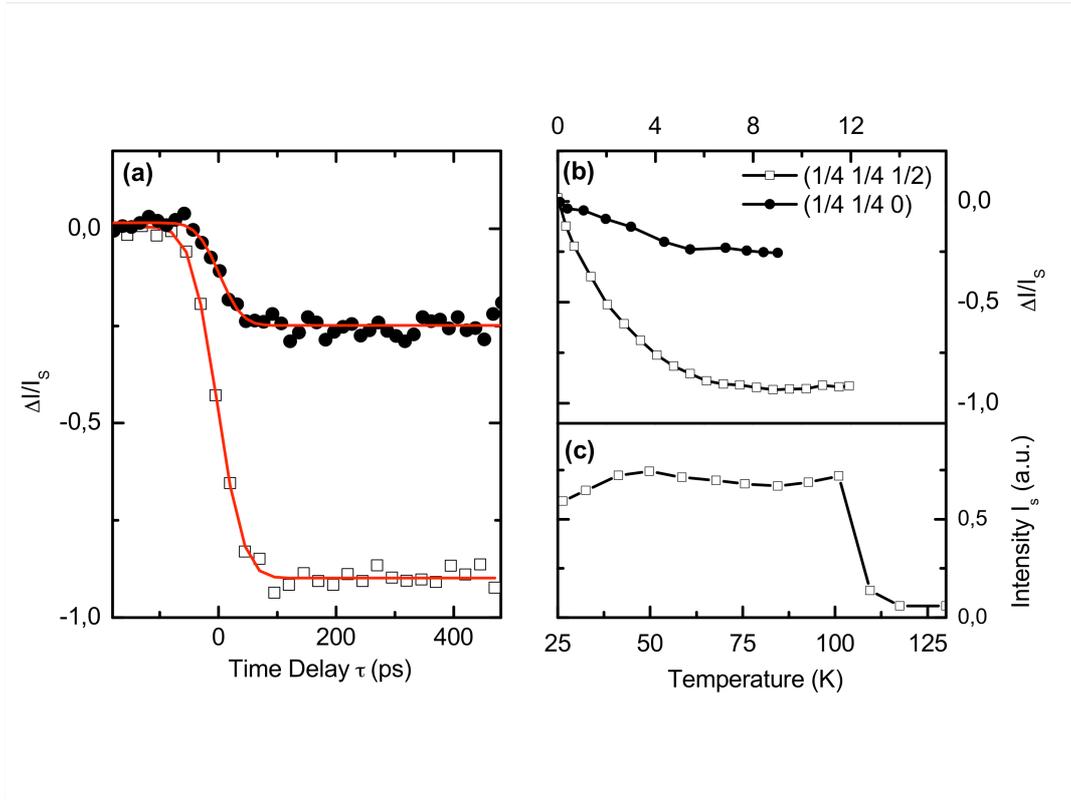

**Figure 2**

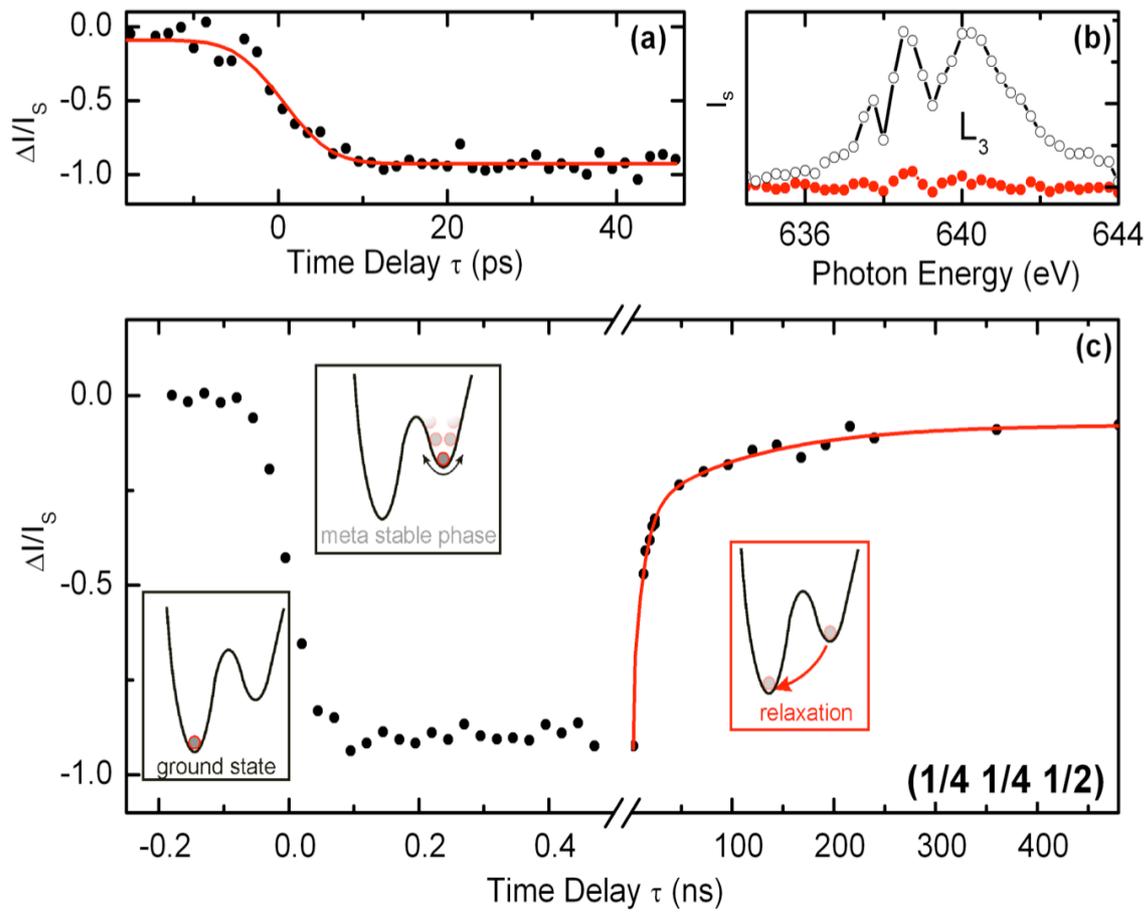

**Figure 3**

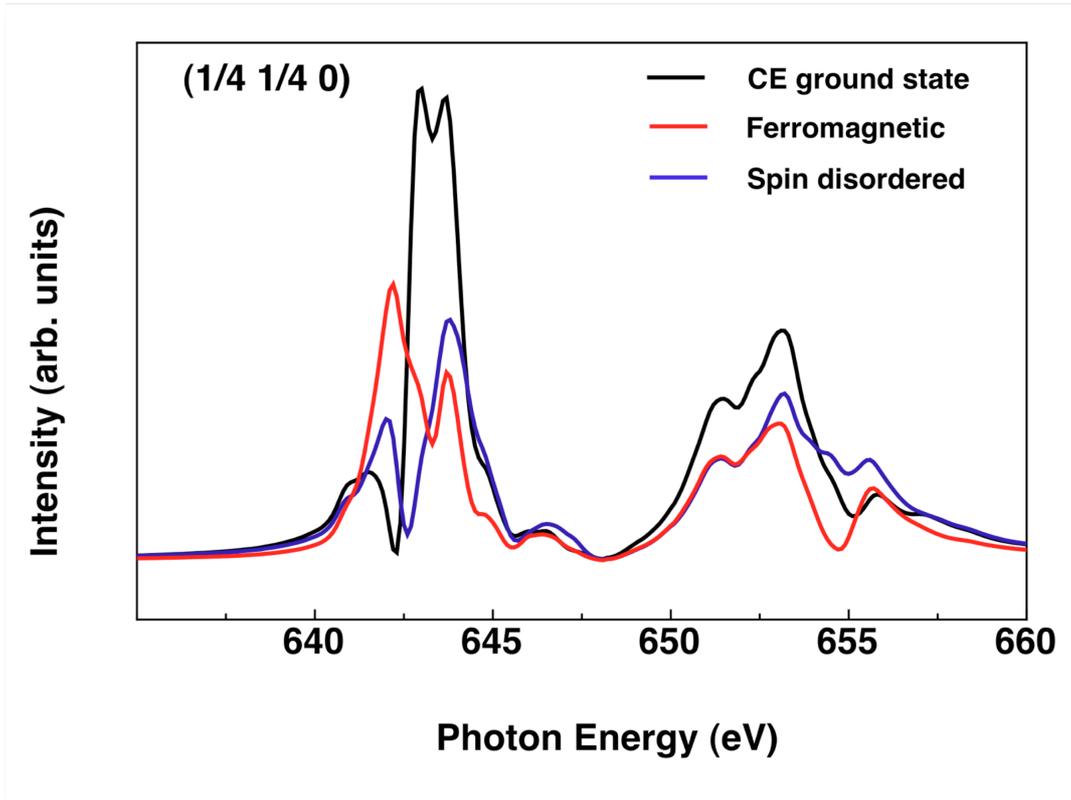

**Figure 4**

**REFERENCES**

*Address correspondence to: dhesi@diamond.ac.uk or andrea.cavalleri@mpsd.cfel.de

[25] We calculate an upper limit for laser-induced heating, reached once equilibrium between the electrons and lattice is established. From optical-conductivity spectra we extract the temperature dependent 800-nm optical absorption depth by Kramers Kroenig transformation, and calculate that a maximum energy density of 72 J/cm$^3$ thermalizes in the 65-nm absorption depth for each mJ/cm$^2$ of incident laser fluence. In the absence of any heat diffusion, which would further decrease the temperature jump, this corresponds to ≤14 K temperature increase for 1 mJ/cm$^2$ at 25 K base temperature, where the heat capacity is about 5 J/(K cm$^3$). As the base temperature increases, higher heat capacities (reaching 19 J/Kcm$^3$ at 100 K) lead to lower heating. The combined temperature increase for 5-mJ/cm$^2$ excitation fluence is ≤60 K. We stress that this is most likely overestimated, since carrier and heat diffusion will in fact transport a fraction of the energy into the bulk before thermalization.

[29] Higher-lying states were also explored, although no notable state was found up to 0.7-eV crystal-field excitations, in which the e$_g$ electron rotated to the *c*-axis and, at higher excess energies, states with a spin-flipped t$_{2g}$ electron replacing the e$_g$ electron on the active Mn$^{3+}$ ion. Whilst these states may be important at the earliest times, we have excluded them from the contributors to the metastable state.